\documentclass{IEEEtran}

\usepackage[hidelinks]{hyperref}
\usepackage{caption}

\usepackage{indentfirst}

\usepackage{xcolor}
\usepackage{graphicx}
\usepackage{epstopdf}
\graphicspath{{fig/}}
\usepackage{subfigure}
\usepackage{float}
\usepackage{array}
\usepackage{url}
\usepackage{longtable}
\usepackage{rotating}
\usepackage{multirow}
\usepackage{amsmath}
\usepackage{amsfonts}
\usepackage{amssymb}
\usepackage{mathrsfs}
\usepackage{mdwlist}
\usepackage{booktabs}
\usepackage{threeparttable}
\usepackage{makecell}
\usepackage{diagbox}
\usepackage{pifont}
\usepackage{xcolor}
\usepackage{booktabs}
\usepackage{adjustbox}

\usepackage{pdflscape}
\usepackage{footnote}

\usepackage{bm}

\usepackage[numbers,sort&compress]{natbib}
\usepackage[linesnumbered,boxed,ruled]{algorithm2e}
\usepackage[hang,flushmargin]{footmisc}
\usepackage{lipsum}

\usepackage[T1]{fontenc}
\usepackage[utf8]{inputenc}

\usepackage{color,xcolor}
\definecolor{myred}{RGB}{217,46,127}
\definecolor{mygreen}{RGB}{67,127,127}

\title{UniForm: A Unified Multi-Task Diffusion Transformer for Audio-Video Generation}

\author{
    {Lei Zhao, Linfeng Feng, Dongxu Ge, Rujin Chen, Fangqiu Yi, Xiao-Lei Zhang, Chi Zhang and Xuelong Li}

    \thanks{{The demo is available at \url{https://uniform-t2av.github.io/}.}}
}

\begin{document}
\maketitle

\begin{abstract}
With the rise of diffusion models, audio-video generation has been revolutionized. However, most existing methods rely on separate modules for each modality, with limited exploration of unified generative architectures. In addition, many are confined to a single task and small-scale datasets. 
{
To overcome these limitations, we introduce UniForm, a unified multi-task diffusion transformer that generates both audio and visual modalities in a shared latent space. 
By using a unified denoising network, UniForm captures the inherent correlations between sound and vision.
Additionally, we propose task-specific noise schemes and task tokens, enabling the model to support multiple tasks with a single set of parameters, including video-to-audio, audio-to-video and text-to-audio-video generation.}
Furthermore, by leveraging large language models and a large-scale text-audio-video combined dataset, UniForm achieves greater generative diversity than prior approaches. {Experiments show that UniForm achieves performance close to the state-of-the-art single-task models across three generation tasks, with generated content that is not only highly aligned with real-world data distributions but also enables more diverse and fine-grained generation.}
\end{abstract}

\begin{IEEEkeywords}
    Text-to-spatial-audio, audio generation, latent diffusion model.
\end{IEEEkeywords}

\section{Introduction}\label{sec:intro}



With the flourishing of deep learning, artificial intelligence generated content (AIGC) has revolutionized multimodal creation, enabling vivid generation across text \cite{chung2024scaling, li2024enhanced}, images \cite{rombach2022high, zhu2023label}, audio \cite{liu2024audioldm2, wang2024continuous}, and video \cite{song2024moviellm, zheng2024open}. This progress has driven innovation in creative industries and expanded the scope of digital media. However, most AIGC systems remain confined to single modality. For example, text-to-video methods, like \cite{singer2023makeavideo} decompose temporal U-Nets into spatial and temporal blocks, \cite{blattmann2023align} extends latent diffusion to video by modeling time, and \cite{wang2024lavie} uses cascaded diffusion with joint image-video tuning. Despite strong visual results, these methods lack sound, overlooking multisensory integration, which is a key element for immersive experiences.

Recent efforts have begun exploring audio-video co-generation. MM-Diffusion \cite{ruan2023mm} employs dual U-Net subnets for parallel audio-video synthesis. Subsequently, MM-LDM \cite{sun2024mm} utilizes two separate diffusion model to independently process audio and video, ultimately enabling multimodal interaction within a shared high-level semantic space.
In contrast, another emerging diffusion backbone is the Diffusion Transformer (DiT) \cite{peebles2023scalable}, which has demonstrated remarkable performance in various content generation tasks. Building on this, \cite{hayakawa2024discriminator} employs two diffusion processes, followed by a joint discriminator to integrate audio and video. Meanwhile, AV-DiT \cite{wang2024av} adopts a shared DiT backbone pre-trained exclusively on image data, facilitating audio and video generation through the addition of lightweight adapters.
\begin{figure}[t]
  \centering
  \includegraphics[width=0.5\textwidth]{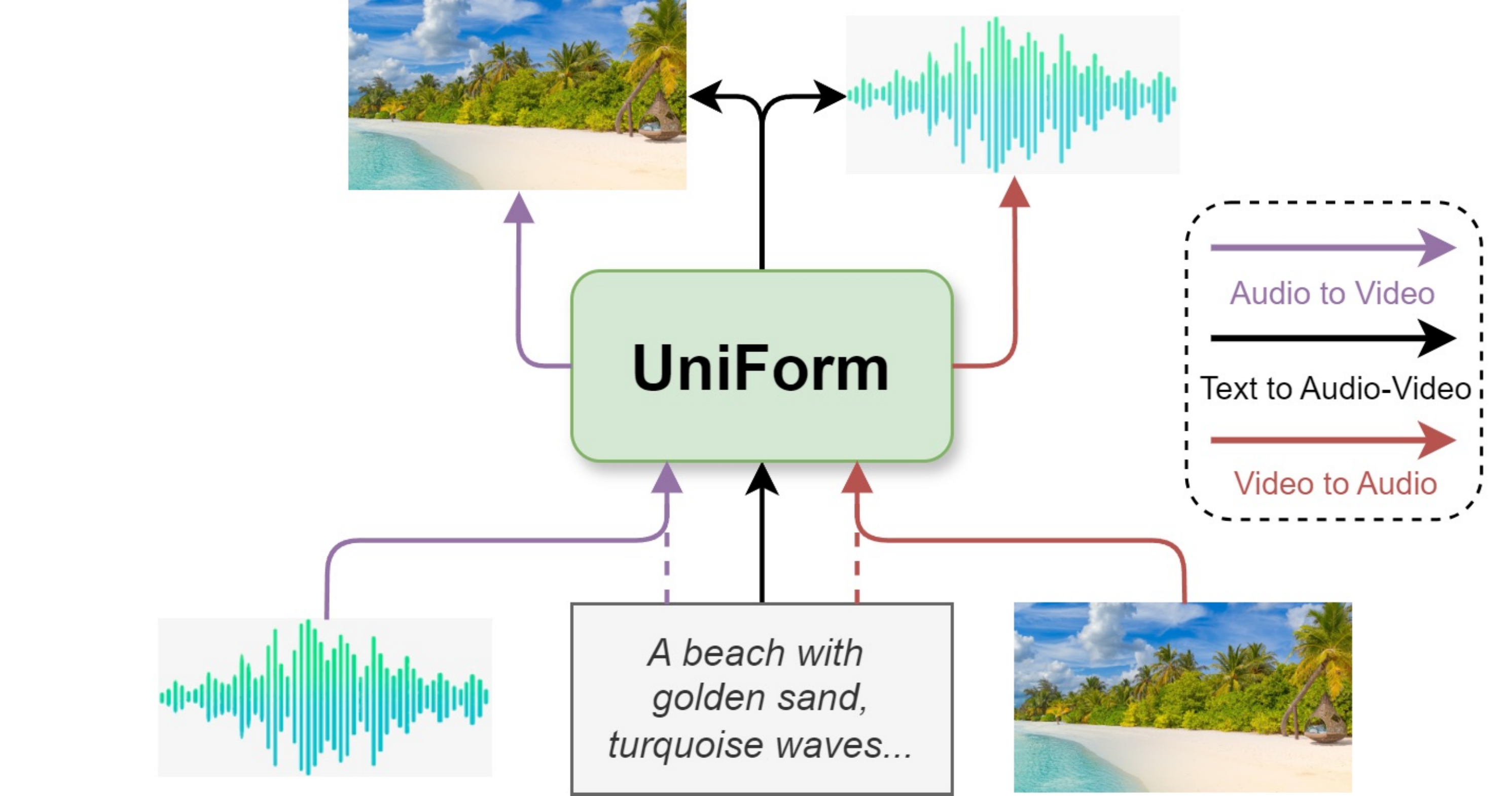}
  \caption{Illustration of multimodal-conditioned audio-video generation. Text can create audio-video directly; audio or video can serve as a condition to guide the generation of the other.}
  \label{fig:illu}
\end{figure}
Although these co-generation methods have shown strong results, there remains room for further exploration. First, they rely on two separate sub-networks to generate audio and video independently, which may limit the depth of cross-modal integration. Second, these methods rely solely on text labels and are mostly trained on small-scale datasets, which limits the diversity of the generated results. 
{Finally, most methods focus solely on the text-to-audio-video (T2AV) task and do not provide adequate exploration and experimentation on other related generation tasks, such as audio-to-video (A2V) or video-to-audio (V2A) generation.}
While \cite{xing2024seeing} enables the three tasks, each task requires distinct pre-trained models in their work.

Inspired by the natural coupling of sound and vision in real-world videos, we ask whether a unified model can improve alignment and consistency across modalities. 
{In this work, we propose Uniform, a unified multi-task diffusion transformer for generating consistent audio-video content.
Moreover, as shown in Figure~\ref{fig:illu}, all three tasks are handled by a unified network with shared weights.}
Our contributions are summarized as follows:
\begin{itemize}
    \item
    {\textbf{We propose a unified DiT to jointly generate audio and video content synchronously.} UniForm constructs a unified latent representation space by concatenating the latent features encoded from the audio Variational Autoencoder (VAE) and the video VAE, enabling joint diffusion modeling across modalities and implicitly capturing the correlations between modalities.
    }
    \item
    {\textbf{We propose a unified denoising network that can perform multi-task audio-video generation with a single set of model parameters, including V2A, A2V and T2AV.}}
    We incorporate task-specific noise schemes and task tokens to specify the target task. For the latter two tasks, text prompts can also be optionally used as auxiliary input to enable fine-grained control and enhance performance.
    \item {\textbf{Our method enables the generation of more diverse audio and video content.}}
    We use a large language model (LLM) to encode the text, a process that does not rely on the text labels used in previous methods, thereby providing finer control for the model and enhancing the diversity of the generated content.
    To fully leverage this advantage , we produced an extensive caption corpus to facilitate model training.
    \item \textbf{Experiments show that UniForm achieves performance comparable to the state-of-the-art single-task baselines.} Remarkably, this performance is achieved without fine-tuning on task-specific datasets, as the model is trained solely in a multi-task setting. In addition, compared to non-unified methods (i.e., using separate backbones for each task \cite{xing2024seeing}), our approach demonstrates consistent advantages across the board.
\end{itemize}

The rest of this paper is organized as follows. 
Section~\ref{sec:relatedworks} discusses related work.
Section \ref{sec:Method} describes the proposed method in detail. 
Section \ref{sec:exper} presents the experimental setup and results. 
Finally, Section \ref{sec:con} concludes the study.

\section{Related Work}
\label{sec:relatedworks}

\subsection{Video to Audio Generation}
In this paper, we focus on ``Foley'' audio,\footnote{\href{https://www.youtube.com/watch?v=UO3N_PRIgX0}{YouTube: The Magic of Making Sound}} which refers to sound effects added during post-production to enrich the auditory experience of multimedia \cite{choi2023foley}, such as the crunch of leaves or the clinking of glass bottles. Earlier AI-based Foley generation methods were conditioned on class labels \cite{liu2021conditional} or text prompts \cite{liu2023audioldm}. Building on this, recent work has expanded video-to-audio generation. SpecVQGAN \cite{iashin2021taming} adopts a transformer that generates high-fidelity spectrograms from video frames using a VQGAN-based codebook and perceptual audio loss. \cite{du2023conditional} requires both silent video and conditional audio to produce candidate tracks, which are filtered using an audio-visual synchronization model. Diff-Foley \cite{luo2024diff} only requires silent video as input. It first aligns audio-visual features through contrastive pretraining (CAVP), then trains a diffusion model on spectrogram latents conditioned on CAVP features. FoleyCrafter \cite{zhang2024foleycrafter} introduces optional text prompts for finer control and incorporates a semantic adapter and temporal controller for improved alignment. V-AURA \cite{viertola2025temporally} proposes an autoregressive model with high-frame-rate visual encoding and cross-modal fusion for fine-grained temporal precision. VATT \cite{liu2024tell} presents a multi-modal system that generates audio from video with optional text prompts. Departing from GANs and diffusion, FRIEREN \cite{wang2024frieren} uses flow matching as its generative backbone. To better preserve temporal structure, it introduces a non-autoregressive vector field estimator without temporal downsampling.

\subsection{Audio to Video Generation}
Given the high information density of video, video-to-audio generation tasks typically treat video as the primary input, with text serving as an auxiliary cue. In contrast, audio to video generation mainly relies on audio alignment. Due to the limited semantic context provided by audio itself, it is difficult for both humans and machines to distinguish (e.g., distinguishing a given audio clip corresponds to climbing stairs or tap dancing). As a result, audio-to-video generation often depends on text or images to supply the missing context, and dedicated efforts in this direction remain relatively limited. \cite{lee2022sound} learns to generate semantically aligned video from audio. It maps audio into the StyleGAN latent space and refines the output using CLIP-based multimodal embeddings to reinforce audio-visual coherence. TPoS \cite{jeong2023power} follows a stage-wise strategy: it first generates an initial frame from a text prompt, then progressively adapts the visuals based on the audio input. \cite{yariv2024diverse} proposes a lightweight adaptor that translates audio features into the input format expected by a pre-trained text-to-video model, enabling audio-driven video generation with minimal changes to the backbone.

\begin{figure*}[t]
    \centering
    \includegraphics[width=0.80\textwidth]{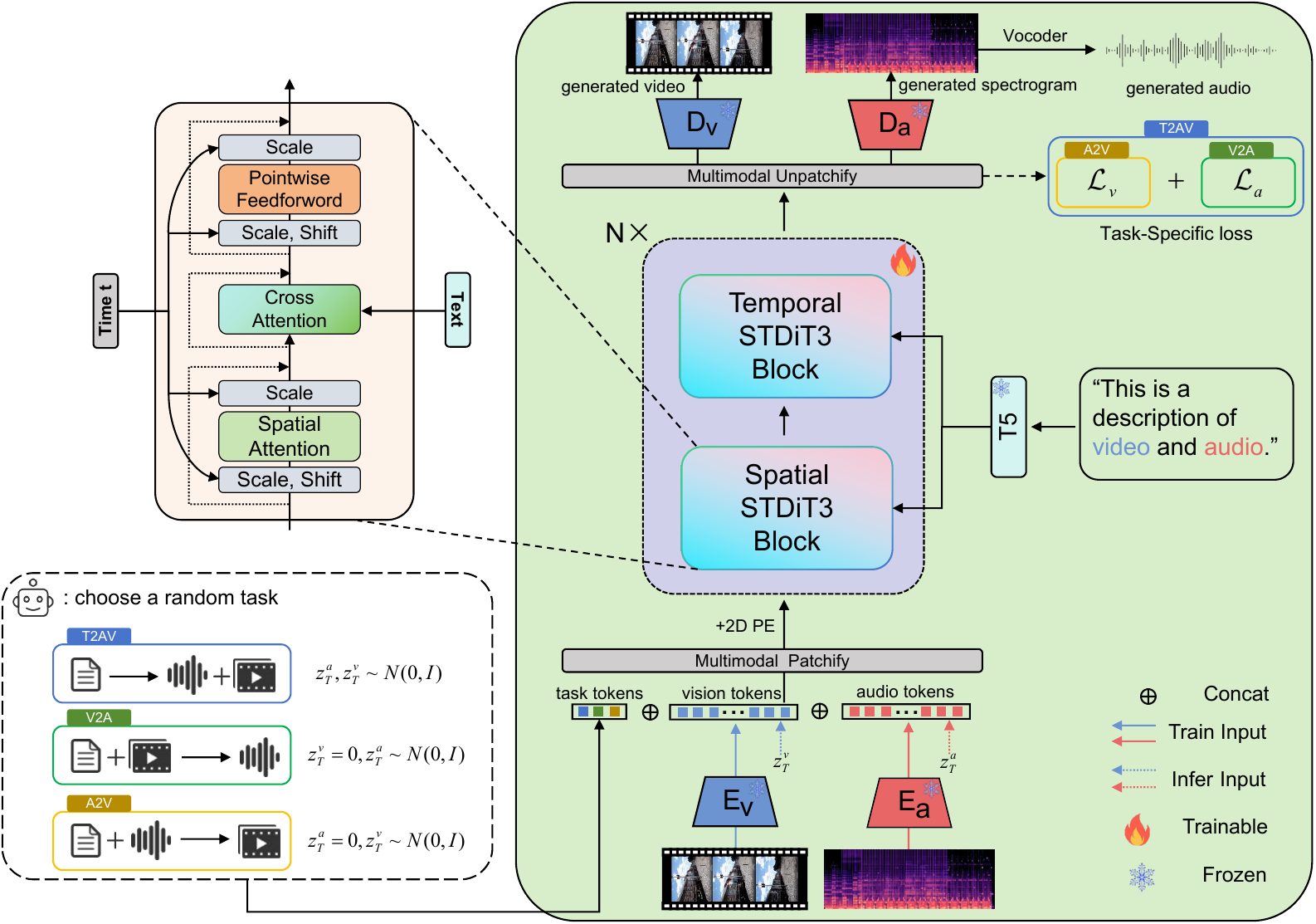}
    \caption{Overview of the proposed UniForm. Vision tokens and audio tokens are integrated and processed within a unified latent space using a DiT model to learn their representations. During training, one of three tasks is randomly selected in each iteration, with task tokens guiding the learning of the DiT. The text encoder, the encoder-decoder for video and audio, and the audio vocoder are all pre-trained models that remain frozen throughout.}
    \label{fig:overview}
\end{figure*}
\subsection{Diffusion-based Generation}\label{subsec:related-diffusion}
Diffusion models, as a class of probabilistic generative models, have received growing attention for their remarkable performance in diverse domains such as image generation \cite{ramesh2022hierarchical} and audio synthesis \cite{liu2024audioldm2}. The majority of existing approaches are built upon Denoising Diffusion Probabilistic Models (DDPMs) \cite{ho2020denoising}, which form the foundation of this paradigm. The key idea of diffusion modeling is to define a forward process that progressively perturbs data into Gaussian noise through a sequence of noise-adding steps. The model is then trained to approximate the reverse process, which starts from pure noise and performs iterative denoising steps to recover samples that approximate the original data distribution. To reduce the computational cost of operating in high-dimensional spaces, Latent Diffusion Models (LDMs) \cite{rombach2022high} shift the diffusion process into a lower-dimensional latent space, enabling more efficient generation. \cite{peebles2023scalable} explore replacing the previously used U-Net backbone with a transformer operating in latent space. Their results show that, given sufficient compute, the Diffusion Transformer (DiT) produces samples that closer to the original data distribution. In this work, we adopt DiT as the backbone of our multi-task architecture, leveraging its scalability and strong performance in modeling a unified latent space across audio and visual modalities.

\section{Method}
\label{sec:Method}

In this section, we first define the three types of audio-video generation tasks addressed in Section \ref{subsec:problem}. Next, we review the preliminary knowledge of diffusion-based generation in Section \ref{subsec:diffusion}. Finally, we present a detailed introduction of the proposed UniForm in Section \ref{subsec:uniform}.

\subsection{Problem Definition}\label{subsec:problem}


Our goal is to enable both audio video to be generated by a single model under varying prior conditions. Here, we define three multimodal generation tasks, including text-to-audio-video (T2AV), audio-to-video (A2V) and video-to-audio (V2A). We denote the denoising network as $\epsilon_\theta$, the text embedding as $c$, the Gaussian noise in the audio and visual modalities as $z^a_T$ and $z^v_T$. The superscripts $a$ and $v$ denote the latent variables for audio and video, respectively. We denote $f$ as the function representing repeated inference using $\epsilon_\theta$ with a sufficient steps. Then, the audio-video generation task can be formulated as follows:
\begin{equation} \label{eq:def}
\hat{z}^a_0, \hat{z}^v_0 = f(z^a_T, z^v_T, c),
\end{equation}

Based on Eq.~\eqref{eq:def}, we introduce three sets of task-specific noise schemes. In the T2AV task, all three inputs are provided. In the A2V task, the audio noise input is removed by setting $z^a_T = 0$; similarly, in the V2A task, we set $z^v_T = 0$. During training, we integrate a classifier-free guidance (CFG) strategy~\cite{ho2021classifier}, conditioning on the text embedding $c$ with a 50\% probability. This allows the model to learn various combinations of modality-specific priors within a single unified framework for multimodal generation.

\subsection{Diffusion Model}\label{subsec:diffusion}
For simplicity without loss of generality, we discuss the case where both the audio and video latent variables are added with noise. The forward diffusion process is defined as a Markovian process from the data distribution to a standard Gaussian distribution by progressively adding noise to the original data sample over discrete time steps $t$. Specifically, noise is incrementally added to the initial true data distributions $z^{a}_{0}$ and $z^{v}_{0}$ through a sequence of Gaussian transitions, governed by a noise schedule $\{\beta_1, \beta_2, \ldots, \beta_t, \ldots, \beta_T\}$ over $T$ total diffusion steps. The forward process is formulated as:
\begin{equation}
q(z^{k}_{t} | z^{k}_{t-1}) = \mathcal{N}\left(\sqrt{1-\beta_{t}} \, z^{k}_{t-1}, \beta_{t} \mathbf{I}\right), k \in \{a,v\},
\end{equation}
\begin{equation}
q(z^{k}_{t} | z^{k}_{0}) = \mathcal{N}\left(\sqrt{\bar{\alpha}_{t}} \, z^{k}_{0}, (1-\bar{\alpha}_{t}) \mathbf{I}\right), k \in \{a, v\},
\end{equation}
where \( \alpha_{t} = 1 - \beta_{t} \), and \( \bar{\alpha}_{t} = \prod_{i=1}^{t} \alpha_{i} \). A reparameterization method~\cite{song2020denoising} simplifies the sampling of any intermediate states \( z^{a}_{n} \) and \( z^{v}_{n} \) from the initial states \( z^{a}_{0} \) and \( z^{v}_{0} \), using the following formulation:
\begin{equation}
z^{k}_{t} = \sqrt{\bar{\alpha}_{t}} \, z^{k}_{0} + \sqrt{1 - \bar{\alpha}_{t}} \, \epsilon^{k}, k \in \{a,v\},
\end{equation}
where \( \epsilon^{m}, \epsilon^{s} \sim \mathcal{N}(\mathbf{0}, \mathbf{I}) \) introduce independent noise. At the final step of forward diffusion, both \( z^{m}_{N} \) and \( z^{s}_{N} \) resemble standard Gaussians.

The goal of the reverse process is to progressively generate \( z^{k}_{0} \)  from $z^k_T, k \in \{a,v\}$. Similar to forward process, reverse process can also be represented as a Markov process. The noise prediction loss $\mathcal{L}$ can be simplified as minimizing the mean square loss between the denoising network prediction and ground-truth added noise in forward process, defined as follows:

\begin{equation}
\mathcal{L} =  \gamma_{t} \sum_{k \in \{a,v\}} \mathbb{E}_{\epsilon^{k}_{t} \sim \mathcal{N}(\mathbf{0}, \mathbf{I}), z^{k}_{0}} \left\| \epsilon^{k}_{t} - \epsilon_{\theta}^{(t)}(z^{k}_{t}, c) \right\|_{2}^{2},
\end{equation}
where $\gamma_{t}$ adjusts the weight of each reverse step based on its signal-to-noise ratio.

Finally, starting from standard Gaussian noises $z^a_T$ and $z^v_T$, the generated data $z^a_0$ and $z^v_0$ can be obtained by progressively sample through:
\begin{equation}
p_{\theta}(z^{k}_{0:T} | c) = p(z^{k}_{T}) \prod_{t=1}^{T} p_{\theta}(z^{k}_{t-1} | z^{k}_{t}, c),
\end{equation}
\begin{equation}
p_{\theta}(z^{k}_{t-1} | z^{k}_{t}, c) = \mathcal{N}\left(\mu_{\theta}^{(t)}(z^{k}_{t}, c), \tilde{\beta}^{(t)}\right),
\end{equation}
\begin{equation}
\mu_{\theta}^{(t)}(z^{k}_{t}, c) = \frac{1}{\sqrt{\alpha_{t}}} \left[ z^{k}_{t} - \frac{1 - \alpha_{t}}{\sqrt{1 - \bar{\alpha}_{t}}} \epsilon_{\theta}^{(t)}(z^{k}_{t}, c) \right],
\end{equation}
\begin{equation}
\tilde{\beta}^{(t)} = \frac{1 - \bar{\alpha}_{t-1}}{1 - \bar{\alpha}_{t}} \beta_{t},
\end{equation}
\[
k \in \{a, v\}.
\]


\subsection{UniForm}\label{subsec:uniform}

\subsubsection{Video \& Audio Latent Encoding}
As mentioned earlier, we adopt DiT to model a unified latent space shared across audio and visual modalities. {FLAN-T5 \cite{colin2020exploring, chung2024scaling} is used as the text encoder, where T5 is a high-capacity pretrained LLM known for its strong semantic understanding.} During training, the video input is represented as $V \in \mathbb{R}^{T^v \times C^v \times H \times W}$, where $T^v$ denotes the number of temporal frames, each with $C^v$ channels, height $H$, and width $W$. A pre-trained video encoder \cite{zheng2024open} is adopted to extract vision latent $z^v_0\in \mathbb{R}^{\hat{C}^v\times \hat{T}^v\times \hat{H}\times\hat{W}}$ from videos, where $\hat{C}^v$, $\hat{T}^v$, $\hat{H}$, $\hat{W}$ are hidden dimensions of vision tokens. For the audio input, we first apply the Short-Time Fourier Transform (STFT) to convert the waveform from the time domain to the frequency domain. Then, a set of Mel-scale filters is used to generate the Mel spectrogram with shape $\mathbb{R}^{T^a \times F}$, where $T^a$ denotes the number of temporal frames and $F$ is the number of frequency bins. These Mel spectrograms are subsequently passed through a pre-trained audio VAE encoder \cite{liu2023audioldm} to obtain the audio latent tokens $z^a_0 \in \mathbb{R}^{\hat{C}^a \times \hat{T}^a \times \hat{F} \times 1}$. Latent tokens from both modalities first undergo reshaping operations to align their dimensions. Subsequently, these adjusted tokens are concatenated along the last dimension, forming a unified representation that serves as the input to the shared DiT.

\subsubsection{Multitask Modeling}
As shown in Figure \ref{fig:overview}, we further incorporate additional task tokens into the input to assist the model in better understanding the task. Specifically, when performing one of the three tasks, the task ID is passed through a task tokenizer to obtain its task token. The task token is linked with latent token via the concatenation operation, thereby forming the final input. Subsequently, the concatenated input is passed through a multimodal patch embedder, which projects the unified latent representation into the suitable embedding space. Additionally, a time embedder is utilized to integrate the diffusion timestamp into the input. 

After obtaining the joint representation of vision as well as audio as input, we adopt STDiT3 \cite{zheng2024open} blocks to progressively integrate information from both spatial and temporal domains. In order to integrate textual information, cross attention mechanism is applied in both STDiT3 blocks, which can be seen in Figure \ref{fig:overview}. Noted that due to spatial limitations, we only presented the spatial version of STDiT3. The temporal version of STDiT3, on the other hand, replaced the spatial attention mechanism within the module with a temporal attention mechanism, while keeping all other settings consistent.


\subsubsection{Video \& Audio Latent Decoding}
Once we obtain the final output from the DiT blocks, we utilize multimodal unpatchify to derive the predicted noise and variance for both the audio and video. During training, the predicted noise is used to compute the loss. During inference, the DiT gradually reduces the noise, ultimately generating latent representation with minimal noise in the final diffusion time step. The latent representation is then divided and reshaped into two distinct modal latent forms, corresponding to the audio and video. Using the decoders of pre-trained VAEs, the latent features of both audio and video are simultaneously reconstructed into generated video frames and audio Mel-spectrograms. Subsequently, these Mel-spectrograms are further converted into audio waveforms using the pre-trained HiFi-GAN \cite{kong2020hifi}.

\subsubsection{Video \& Audio Generation Loss}
Here, we outline the objective of our model in the denoising process for the three tasks that we previously proposed. During V2A task, the audio generation loss $\mathcal{L}_a$ can be formulated as:
\begin{equation}
\mathcal{L}_a =  \left\|\epsilon^{a}_{t} - Mask_a(\epsilon_\theta(z^a_t,z^v_t, c))\right\|_{2}^{2},
\end{equation}
where $Mask_a$ is used to extract the audio token part from the combined noise representation. For the video generation loss $\mathcal{L}_v$ in A2V task, it similarly can be denoted as:
\begin{equation}
\mathcal{L}_v =  \left\|\epsilon^{v}_{t} - Mask_v(\epsilon_\theta(z^a_t,z^v_t, c))\right\|_{2}^{2},
\end{equation}
where $Mask_v$ also represents masking the noise representations except for vision tokens.
As for the T2AV task, the multimodal loss $\mathcal{L}_{av}$ is defined as
\begin{equation}
\mathcal{L}_{av} =  \mathcal{L}_a+\mathcal{L}_v,
\end{equation}
Note that for all three tasks, we utilize CFG scheme, which randomly discards text guidance with a 50\% chance. This approach ensures that our model can sustain its generation performance even without a provided video (or audio) description. 

\section{Experiments}
\label{sec:exper}

\subsection{Experimental Setup}
\subsubsection{Datasets}
The training datasets used in this work include  VGGSound \cite{chen2020vggsound}, Landscape \cite{lee2022sound}, AIST++ \cite{li2021ai}, AudioSet-balance \cite{gemmeke2017audio} and AudioSet-Strong \cite{hershey2021benefit}. 
VGGSound is an extensive single-label audio-visual dataset comprising more than 200,000 videos for 310 audio classes. It is used for various tasks such as audio classification, multi-modal classification, and zero-shot audio classification. 
Landscape is a high-fidelity dataset that encompasses video and audio streams, highlighting nine varied natural scenes, including but not limited to raining, splashing water, thunder, and underwater bubbling.
AIST++ is a dedicated subset constructed from the AIST dance dataset \cite{tsuchida2019aist}, containing 1,020 dance motion sequences spanning 10 distinct dance genres, with a total duration of 5.2 hours. To enhance visual presentation, all videos undergo standardized processing through center-cropping techniques, being uniformly resized to a resolution of 1024×1024 pixels.
AudioSet-balance consists of 22,176 segments selected from the AudioSet \cite{gemmeke2017audio}, with each class having at least 59 samples. AudioSet-Strong is another subset of AudioSet, which involves approximately 67,000 segments with frame-level annotations (with a resolution of 0.1 seconds). We adhered to the settings in \cite{hayakawa2024discriminator} for the training split.

Due to the differences in datasets used across various tasks, we evaluate the proposed model only on the commonly used standard evaluation datasets for each task as detailed in the results section. Specifically, for the T2AV task, we evaluate on the Landscape and AIST++ datasets; for the V2A task, we use the VGGSound dataset for evaluation; and for the A2V task, we use the Landscape dataset for evaluation. To obtain more experimental results on each dataset, please refer to Section A of the supplementary materials.

\begin{table*}[t]
    \centering
    \caption{Comparison of different methods for V2A task.}
    \setlength{\tabcolsep}{7.1mm}
    \begin{tabular}{ccccccc}
        \toprule
        \textbf{Dataset} & \textbf{Method}  & \textbf{FAD}$\downarrow$ & \textbf{FD}$\downarrow$ & \textbf{IS}$\uparrow$ & \textbf{KL}$\downarrow$ & \textbf{AV-align}$\uparrow$ \\
        \midrule
        \multirow{7}{*}{VGGSound}           
        & SpecVQGAN \cite{iashin2021taming} &5.42 &31.69 &5.23 &3.37 & 0.417\\
        & Diff-Foley \cite{luo2024diff} &4.72 &23.94 &11.11 &3.38 & 0.386\\
        & V-AURA \cite{viertola2025temporally} &2.88 &14.80 &10.08 &2.42 & 0.366\\
        & VATT \cite{liu2024tell} &2.77 &10.63 &11.90 &\textbf{1.48} & -\\
        & Frieren \cite{wang2024frieren} & 1.34 & 11.45 & 12.25 & 2.73 & 0.422\\
        & FoleyCrafter \cite{zhang2024foleycrafter} &2.51 &16.24 & \textbf{15.68} &2.30 &0.403\\
        & UniForm (ours) &\textbf{1.30} &\textbf{6.21} &15.43 &2.46 & \textbf{0.430}\\
        \bottomrule
    \end{tabular}
    \label{tab:comparison v2a}
\end{table*}

\begin{table*}[ht]
    \centering
    \caption{Comparison of different methods for A2V task.}
    \setlength{\tabcolsep}{9mm}
    \begin{tabular}{ccccccc}
    \toprule
    \textbf{Dataset} & \textbf{Method}  & \textbf{FVD}$\downarrow$ & \textbf{IS}$\uparrow$ & \textbf{AV-align}$\uparrow$\\
    \midrule
    \multirow{5}{*}{Landscape}
                  & MM-Diffusion \cite{ruan2023mm}                               & 922           & 2.85             & 0.410       \\
                  & TempoToken \cite{yariv2024diverse}                               & 784           & 4.49             & \textbf{0.540}       \\
                  & Sound-guided Video Generation \cite{lee2022sound}            & 544           & 1.16             & -          \\
                  & TPoS \cite{jeong2023power}                                     & 421           & 1.49             & -      \\
                  & UniForm (ours)                                       & \textbf{219}  & \textbf{4.61}        & 0.497    \\
    \bottomrule
    \end{tabular}
\label{tab:comparison a2v}
\end{table*}

\subsubsection{Implementation}
For data preprocessing, we resample the videos to 17 fps and then resize them to a resolution of $256\times 256$, and we resample the audios at 16 kHz. Then, we truncate the first 4s of the video and audio samples as the input for VAEs. The pre-trained VAEs from Open-Sora \cite{zheng2024open} and AudioLDM \cite{liu2023audioldm} are used to encode/decode videos and audios, respectively. We use pLLaVA \cite{xu2024pllava}, a Visual Language Model (VLM), to generate text descriptions \footnote{{Given that audio inherently has a lower information density than video, this limitation constrains the capability of existing audio language models to accurately characterize audio content. Additionally, inconsistent audio-visual descriptions may hinder the model's ability to effectively learn audiovisual synchronization. Notably, detailed video captions often inherently encompass attributes of sound-producing objects, including target features, motion patterns, and intensity levels. For these reasons, we exclusively adopt video captions as textual conditions without incorporating audio captions.}}. As for the Landscape and AIST++ dataset, we use its class labels as captions. 
Unless otherwise stated, the subsequent sections assume text use as the default condition.
Our DiT model utilizes pre-trained weights from image generation \cite{chen2024pixart}. We set the batch size to 32 and conducted 230 epochs of iterations with a constant learning rate of $1\times10^{-4}$ using the HybirdAdam optimizer. Linear warmup is adopted as the learning rate scheduling strategy, and the warmup step is set to 1000. In the inference stage, the number of inference steps for each task is uniformly set to 30, and the value of classifier-free guidance is set to 5.

 \begin{figure}[t]
    \centering
    \includegraphics[width=0.46\textwidth]{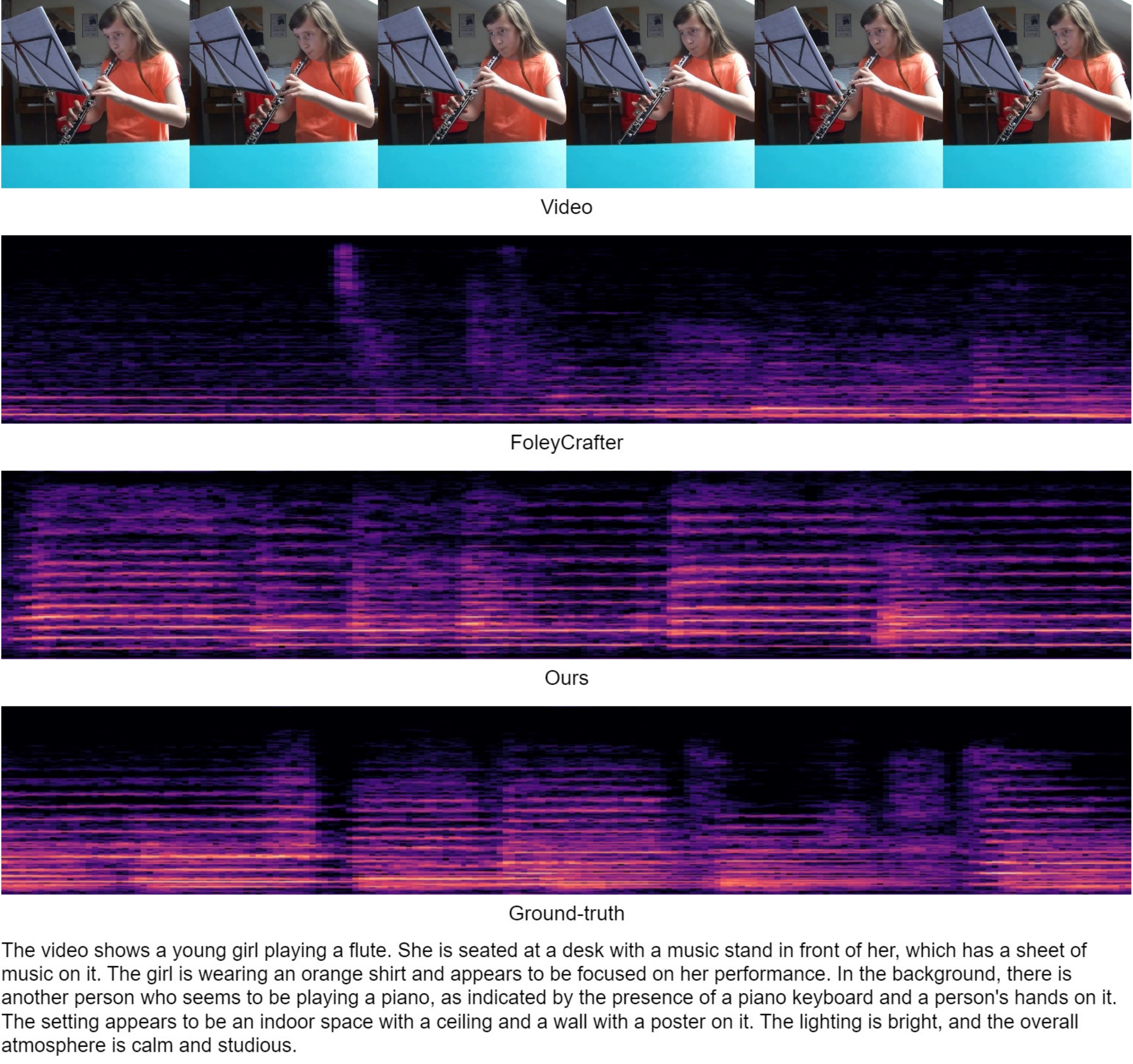}
    \caption{Compared with FoleyCrafter in V2A generation on the VGGSound dataset. Our method can generate more accurate prosody and richer high-frequency details.}
    \label{fig:tv2a}
\end{figure}

\begin{figure*}[t]
    \centering
    \includegraphics[width=0.96\textwidth]{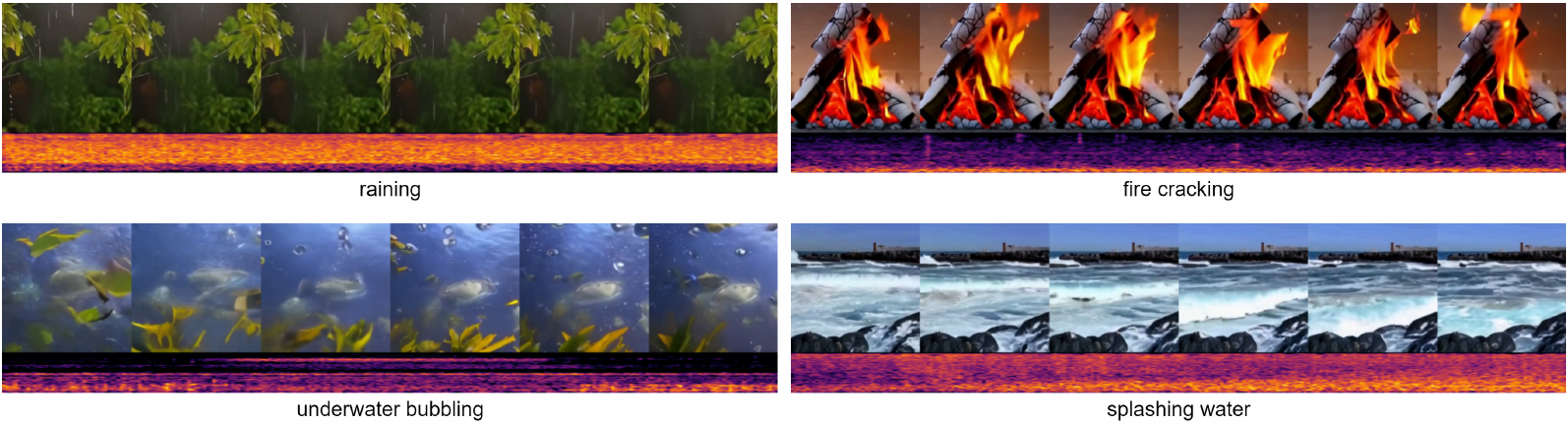}
    \caption{Generated samples in the A2V task on the Landscape dataset.}
    \label{fig:ta2v_demo}
\end{figure*}

\subsubsection{Evaluation metrics}
For evaluating video generation, we adpot the Frechet Video Distance (FVD), Kernel Video Distance (KVD) and Inception Score (IS). 
For audio generation, our evaluation relies on the metrics Frechet Audio Distance (FAD), Frechet Distance (FD), kullback–leibler divergence (KL) and IS. Additionally, we utilize the AV-align \cite{xing2024seeing} metric to assess the synchronization between generated audio and video.

\subsection{Results on Video to Audio Generation}

Table \ref{tab:comparison v2a} lists the comparison results of UniForm with some recent single-task methods on V2A generation, with the evaluation dataset being the VGGSound test set.
As can be observed from the table, our approach outperforms most baselines across the majority of metrics. Specifically, it achieved the top rankings in the FAD and FD metrics with scores of 1.3 and 6.21, respectively, surpassing Frieren \cite{wang2024frieren} and VATT \cite{liu2024tell}, which ranked second in their respective categories. The two methods achieved FAD scores of 1.34 and 2.77, and FD scores of 11.45 and 10.63, respectively. Additionally, our model secured the second position in the IS metric with a score of 15.43, closely followed by FoleyCrafter \cite{zhang2024foleycrafter}, which ranked first with an IS score of 15.68. Despite exhibiting merely average performance in KL, UniForm achieved the highest score in the AV-align metric. This result demonstrates that our approach effectively enhances the alignment between video and audio. It is noteworthy that other baseline methods are only applicable to the task of V2A. This highlights that our multi-task model can generate audio that is highly relevant to the video content, with a quality that rivals the best current V2A approaches.

Figure \ref{fig:tv2a} presents a visual comparison between the mel spectrograms of the audio generated by our method and that produced by FoleyCrafter \cite{zhang2024foleycrafter}. Compared to FoleyCrafter, our method exhibits higher visual correlation with the ground truth, especially noticeable in the first half of the mel spectrogram where FoleyCrafter lacks certain elements. Additionally, our approach captures more high-frequency details.

\begin{table*}[ht]
    \caption{Comparison of different methods for T2AV task.}
    \centering
    \setlength{\tabcolsep}{4.4mm}
    \begin{tabular}{c|cccc|cccc}
        \toprule
        & \multicolumn{4}{c}{\textbf{Landscape}} & \multicolumn{4}{c}{\textbf{AIST++}} \\\hline
        \textbf{Method}  & \textbf{FAD}$\downarrow$&  \textbf{FVD}$\downarrow$ &\textbf{KVD}$\downarrow$ & \textbf{AV-align}$\uparrow$ & \textbf{FAD}$\downarrow$  &\textbf{FVD}$\downarrow$ & \textbf{KVD}$\downarrow$ & \textbf{AV-align}$\uparrow$\\
        \hline
          MM-Diffusion \cite{ruan2023mm} & 10.61 & 186  & 9.21  & 0.261 &10.58 & 98 & 18.90 &0.273\\
          {MM-LDM} \cite{sun2024mm} & 9.1 & \textbf{77}  & \textbf{3.20} & - & 10.2 & \textbf{55} & \textbf{8.20}  &-\\   
          AV-DiT \cite{wang2024av} & 11.17 & 172  & 15.41 & - &10.17 & {68} & 21.01  &-\\
          MMDisCo \cite{hayakawa2024discriminator} & 5.52 & 405  & - & - &2.17 & 450 & -  &-\\       
          UniForm (ours) & \textbf{2.41} & 164 & 9.05 & \textbf{0.305} & \textbf{1.27}  & 62 & 17.24 & \textbf{0.283}\\
        \bottomrule
    \end{tabular}
    \label{tab:comparison t2av}
\end{table*}

\begin{figure*}[t]
    \centering
    \includegraphics[width=0.96\textwidth]{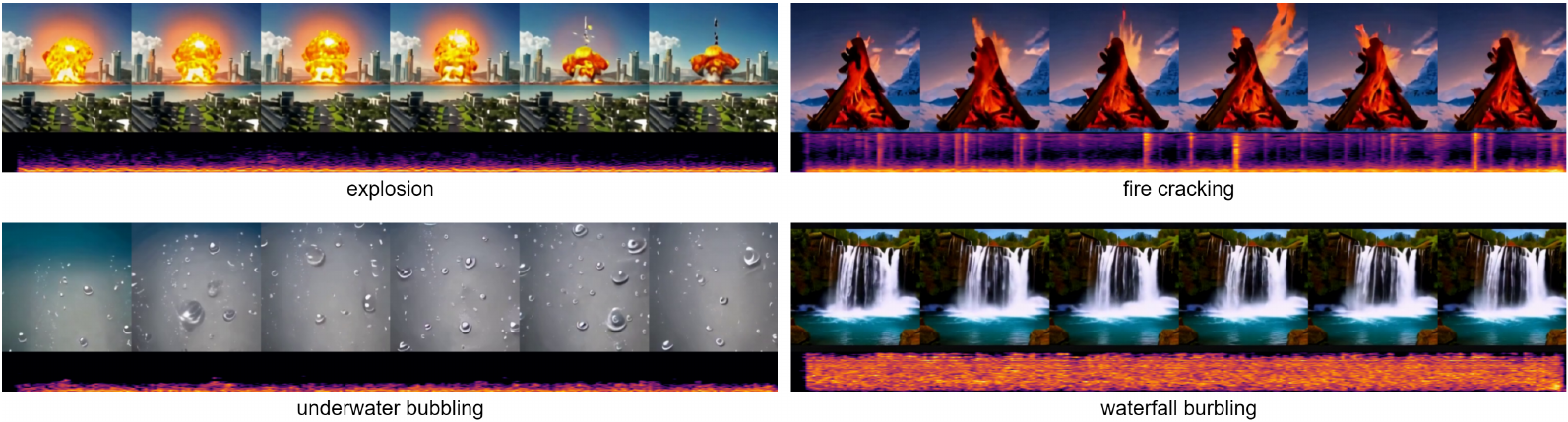}
    \caption{Generated samples in the T2AV task on the Landscape dataset.}
    \label{fig:t2av_demo}
\end{figure*}

\begin{figure}[t]
    \centering
    \includegraphics[width=0.46\textwidth]{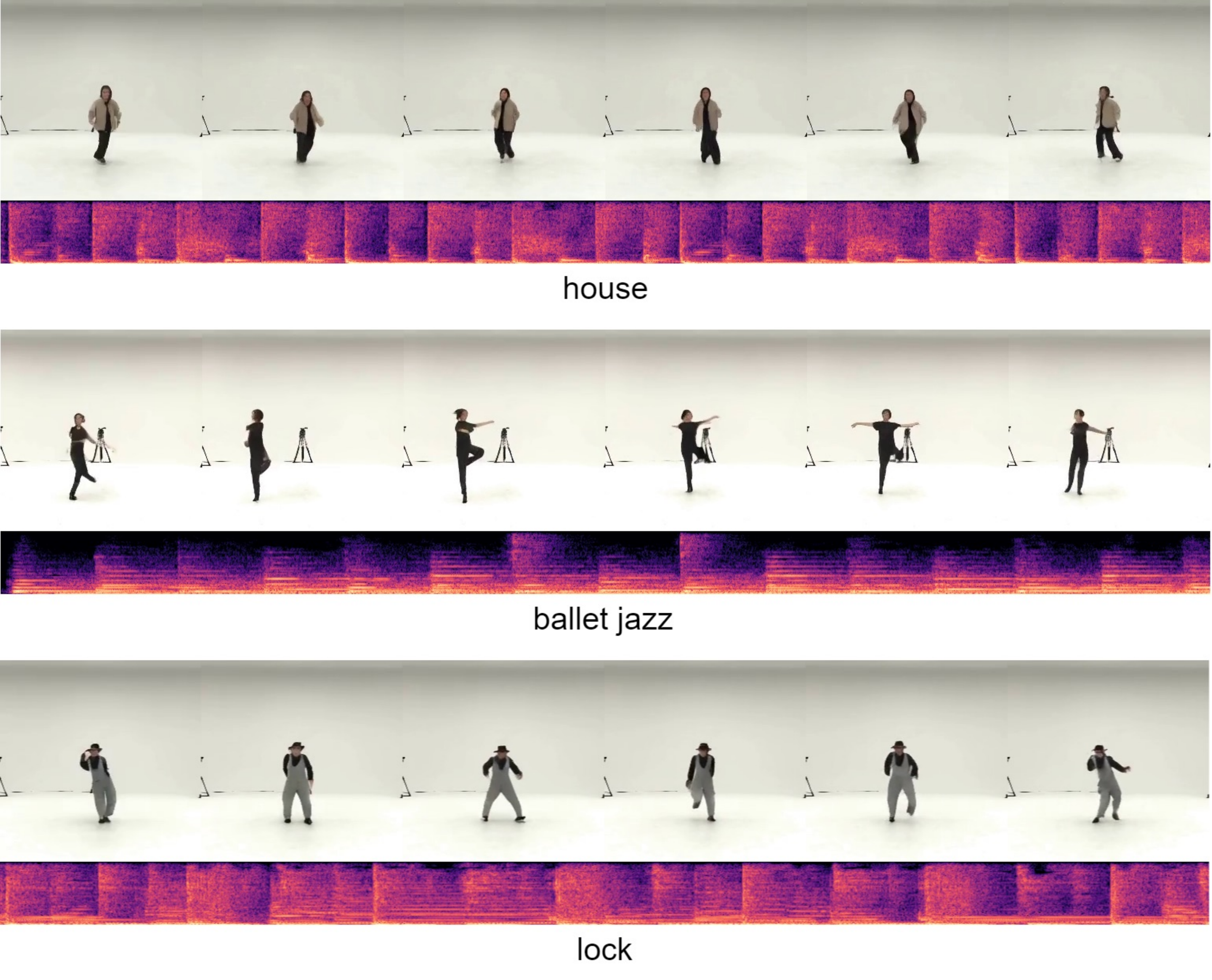}
    \caption{Generated samples in the T2AV task on the AIST++ dataset.}
    \label{fig:t2av_aist}
\end{figure}

\begin{figure*}[t]
    \centering
    \includegraphics[width=0.96\textwidth]{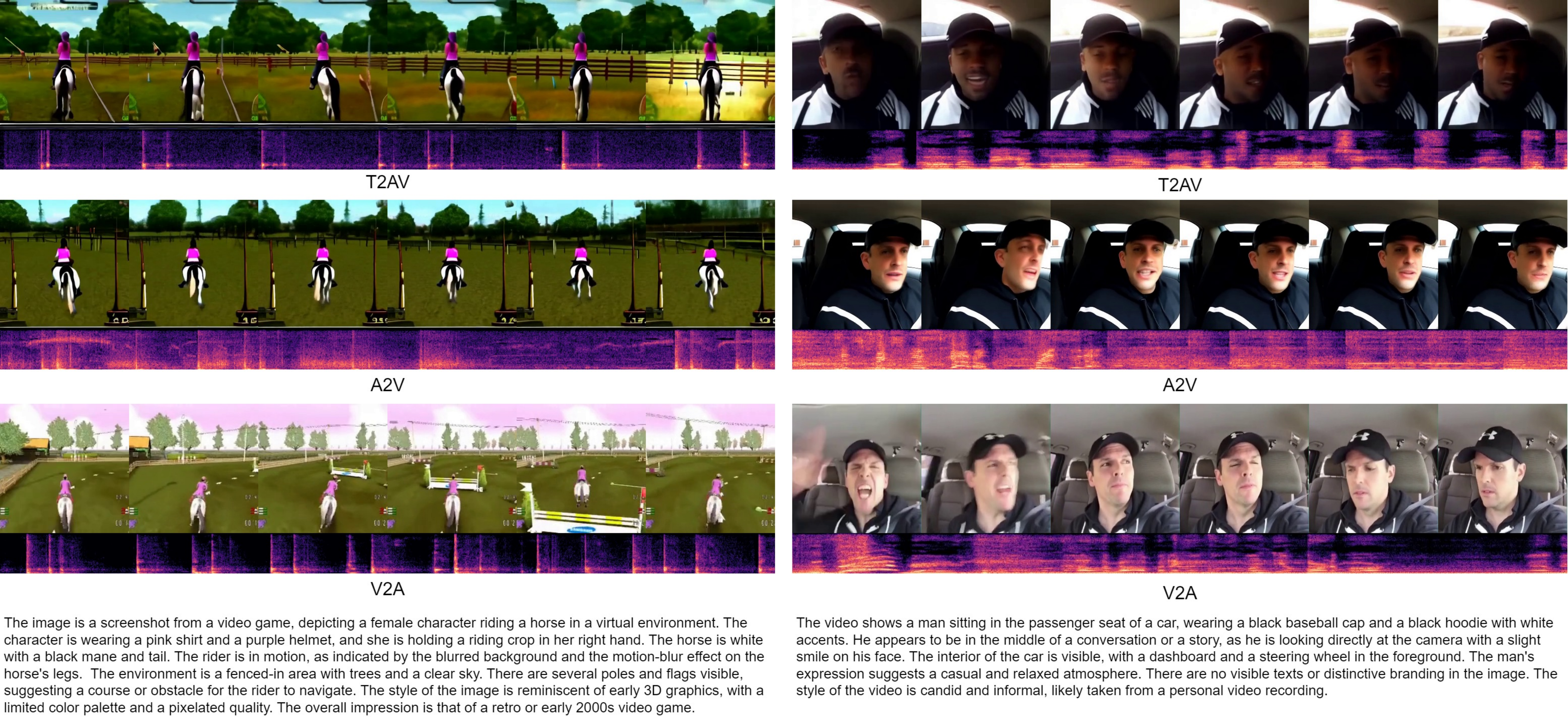}
    \caption{Generated two challenging samples on the VGGSound dataset.}
    \label{fig:multi_task_demo}
\end{figure*}
\subsection{Results on Audio to Video Generation}
Table \ref{tab:comparison a2v} presents the comparison between UniForm and A2V-specific baselines on the Landscape dataset.
Specifically, our approach achieves the lowest FVD score of 219, demonstrating its ability to generate videos with the highest quality. Meanwhile, in terms of IS, which measures content diversity and quality, our approach also leads, achieving a score of 4.61. This score showcases the strength of our method in generating varied content. Although our method attains an AV-align score of 0.497, slightly lower than the top-ranked TempoToken's 0.54, our approach overall excels in audio-to-video generation. It proves that UniForm can significantly improve the quality of generated videos while enhancing audio-visual synchronization. Figure \ref{fig:ta2v_demo} presents four representative examples generated by our method on the Landscape dataset.

\subsection{Results on Joint Audio-Video Generation}

Table \ref{tab:comparison t2av} lists the comparison results of  different methods for joint video and audio generation on the Landscape dataset and AIST++ dataset. 
{In evaluations on the Landscape and AIST++ datasets, the UniForm algorithm outperforms MM-Diffusion, AV-DiT, and MMDisCo across various metrics for both audio and video generation.
Compared to MM-LDM, our method performs worse on video generation metrics across the two datasets but shows significant superiority in audio generation. The gap in video generation may be due to MM-LDM being trained separately on the two datasets with dataset-specific training (unlike our approach based on large-scale data training), resulting in generated distributions that are more closely aligned with the original data distributions.
Additionally, our method also outperforms the comparative methods in audio-video consistency.} 
Figures \ref{fig:t2av_demo} and \ref{fig:t2av_aist} showcase some generation examples of our method on the Landscape dataset and AIST++ dataset, respectively.

\begin{table*}[t]
\centering
\caption{Comparison with non-unified method across three tasks.}
\setlength{\tabcolsep}{6.65mm}
\label{tab:seeing}
{
\begin{tabular}{ccccccc}
\toprule
\textbf{TASK} & \textbf{Model} & \multicolumn{1}{c}{\textbf{FAD}$\downarrow$} & \multicolumn{1}{c}{\textbf{FD}$\downarrow$} & \multicolumn{1}{c}{\textbf{IS}$\uparrow$} & \multicolumn{1}{c}{\textbf{KL}$\downarrow$} & \textbf{AV-align}$\uparrow$\\ \midrule
\multirow{2}{*}{\textbf{V2A}} & Seeing\&Hearing & 5.40 & 24.58 & 8.58 & \textbf{2.26} & 0.411\\
                     & UniForm (ours)  & \textbf{1.30} & \textbf{6.21} & \textbf{15.43} &2.46 & \textbf{0.430}\\
\midrule &             
& \textbf{FVD}$\downarrow$ & \textbf{KVD}$\downarrow$& \textbf{IS}$\uparrow$ & \textbf{AV-align}$\uparrow$ & \\
\midrule
\multirow{2}{*}{\textbf{A2V}} & Seeing\&Hearing  & 402 & 34.76 & - & \textbf{0.522} & \\
              & UniForm (ours) &  \textbf{92} & \textbf{8.05}& 9.50  & 0.483 & \\ 
\midrule &             
& \textbf{FAD}$\downarrow$ &\textbf{FVD}$\downarrow$& \textbf{KVD}$\downarrow$ & \textbf{AV-align}$\uparrow$ &\\
\midrule
\multirow{2}{*}{\textbf{T2AV}} & Seeing\&Hearing  & 12.76 & 326 & \textbf{9.20} & 0.283&\\
              & UniForm (ours) & \textbf{2.41} & \textbf{164}  & 9.05 & \textbf{0.305}&\\ 

\bottomrule
\end{tabular}
}
\end{table*}
Figure \ref{fig:multi_task_demo} displays two challenging generation examples on three tasks. In the left figure, we present the generation results of a game scene. Despite a significant distribution shift between the scenario and the real-world environment, UniForm demonstrates remarkable generation capability. The right-side illustration demonstrates speech-synchronized portrait instances generated by UniForm. It should be noted that due to the inherent limitations of current VLM-based automatic captioning techniques, discrepancies exist between the generated textual description and the original video content. As can be observed, the generated results exhibit a high degree of consistency with the corresponding text. The generative capability demonstrated in this figure surpasses previous T2AV approaches, which were trained solely on toy datasets and consequently lack comparable efficacy.

\subsection{Comparison with Non-unified Method}
Table \ref{tab:seeing} compares the performance of the UniForm method with non-uniform methods, namely Seeing\&Hearing \cite{xing2024seeing}, across various tasks. Seeing\&Hearing employs corresponding distinct pre-trained models when handling different tasks. We evaluate models' performance on the VGGSound dataset for the V2A and A2V tasks, while for the T2AV task, the Landscape dataset was adopted. This is consistent with the eval datasets used by Seeing\&Hearing. It can be observed that our method surpasses the comparative approach across all metrics and tasks, with the exceptions of KL in the V2A task, AV-align in the A2V task, and KVD in the T2AV task. The results highlight the superiority of adopting a unified backbone approach.

\begin{table*}[t]
\centering
\caption{The influence of text on V2A and A2V tasks.}
    \setlength{\tabcolsep}{7.6mm}
\label{tab:ablation}
\begin{tabular}{ccccccc}
\toprule
\textbf{TASK} & \textbf{Use text?} & \multicolumn{1}{c}{\textbf{FAD}$\downarrow$} & \multicolumn{1}{c}{\textbf{FD}$\downarrow$} & \multicolumn{1}{c}{\textbf{IS}$\uparrow$} & \multicolumn{1}{c}{\textbf{KL}$\downarrow$} & \textbf{AV-align}$\uparrow$\\ \midrule
\multirow{2}{*}{V2A} & \ding{55}  & 2.93 & 12.66 & 6.26 & 3.14 & \textbf{0.433}\\
                     & \ding{51}  & \textbf{1.30} & \textbf{6.21} & \textbf{15.43} &\textbf{2.46} & 0.430\\
\midrule &             
& \textbf{FVD}$\downarrow$ & \textbf{FVD}$\downarrow$& \textbf{IS}$\uparrow$ & \textbf{AV-align}$\uparrow$ &\\
\midrule
\multirow{2}{*}{A2V} & \ding{55} & 545 & 42.96& 3.01 & 0.329 &\\
              & \ding{51} & \textbf{219} & \textbf{14.31}& \textbf{4.61}  & \textbf{0.497} &\\ 
\bottomrule
\end{tabular}
\end{table*}

\begin{table}[t]
\centering
\caption{The Improvement of Alignment in Audio-Visual Joint Generation Compared to Single-modal Generation.}
\setlength{\tabcolsep}{4.2mm}
\begin{tabular}{cccc}
\toprule
 \multirow{2}{*}{\textbf{Method}} & \multicolumn{3}{c}{\textbf{AV-align}} \\
\cline{2-4}
  & \textbf{VGGsound} & \textbf{Landscape} & \textbf{AIST++} \\
\midrule
Unimodal & 0.303 & 0.260 & 0.265 \\
UniForm (Ours) & \textbf{0.374} & \textbf{0.305} & \textbf{0.283} \\
\bottomrule
\end{tabular}
\label{tab:ablation_align}
\end{table}

\subsection{Ablations}

\subsubsection{The Impact of Text Prompts}
Table \ref{tab:ablation} demonstrates the influence of text on both V2A and A2V tasks. The V2A task was evaluated on the landscape dataset, while the A2V task was conducted on the VGGSound dataset. According to the table, using text prompts can effectively improve all metrics of all tasks, except for the AV-align in the V2A task, where the scores before and after using text prompts are almost identical. Overall, introducing text conditions has noticeably improved the generation performance of both tasks.

\subsubsection{The Improvement of Alignment in Audio-Visual Joint Generation Compared to Single-modal Generation}
Table \ref{tab:ablation_align} demonstrates the improvement in alignment on three datasets achieved by the audio-visual joint generative model compared to unimodal generative models, i.e., Unimodal. Unimodal refers to a model where audio and video are independently generated under text guidance, sharing the same architecture as UniForm. As shown in the table, UniForm achieves higher AV-align scores than the independent audio and video generation models across all three datasets. This demonstrates that the proposed model significantly enhances alignment performance between audio and visual modalities, further underscoring the importance of employing a joint generation strategy.

\section{Conclusions}
\label{sec:con}
We have introduced a novel unified multi-task audio-video generation model, UniForm, which achieves simultaneous audio and video synthesis using a single diffusion framework. Built on a diffusion transformer backbone, it employs distinct task tokens to enable audio-video synthesis under varying conditions. It simultaneously supports three generation tasks: text-to-audio-video, audio-to-video, and video-to-audio. Our approach enhances both the generation quality of audio and video and their multimodal alignment. UniForm achieves state-of-the-art generation quality, as demonstrated by subjective perception and objective metric evaluations. This performance is attained without the need for task-specific fine-tuning.

\small
\bibliographystyle{IEEEtran}
\bibliography{Reference}

\end{document}